\documentclass[%
 reprint,
 amsmath,amssymb,
 aps,
pra,
]{revtex4-1}

\usepackage{graphicx}
\usepackage{subfigure}
\usepackage{overpic}
\usepackage{dcolumn}
\usepackage{bm}
\usepackage{color}

\usepackage{soul}
\setstcolor{red}

\begin{document}

\preprint{APS/123-QED}

\title{Single Particle Spectroscopies of $p$-wave and $d$-wave Interacting Bose Gases in Normal Phase}

\author{Zeqing Wang}
\author{Ran Qi}%
 \email{qiran@ruc.edu.cn}
\affiliation{%
 Department of Physics, Renmin University of China, Beijing, 100872, P. R. China
 \\}%

\date{\today}

\begin{abstract}
Motivated by experiments of interacting quantum gases across high partial wave resonance, we
investigated the thermodynamic properties as well as single particle spectrums of Bose gases in
normal phase for different interaction strengths both for $p$-wave and $d$-wave interactions. The equation of state, contact density, momentum distributions, and self-energies of single particle Green's functions are obtained in the spirit of ladder diagram approximations. Radio frequency (RF) spectrum, as an important experimental
approach to detect Feshbach molecules or interaction effect, is calculated at different
temperatures. A reversed temperature dependence on the BEC side and BCS side is identified
for both $p$-wave and $d$-wave interactions. An estimation for the signal of RF spectra under typical
experimental conditions is also provided.

\end{abstract}

\maketitle


\section{Introduction}
Ultracold atomic gases, due to their tunable interaction strength by Feshbach resonance (FR), provide an ideal platform to realize different kinds of quantum phases \cite{Bloch_2008, Giorgini_2008, Chin_2010}.
In contrast to the widely studied $s$-wave FRs, the high partial wave FRs with nonzero relative angular momentum support various kinds of topological quantum matters such as topological superfluids and majorana zero modes.
In recent years, interacting quantum gases across $p$-wave and $d$-wave resonance have been realized and used to investigate the few-body and many-body problems in many experiments \cite{luciuk_evidence_2016, XPLiu_2018, MW_2018, venu_unitary_p_2023, GM_2019, YTChang_2020, DJM_2021, GuoqiB_2022, DM_2023, cui_2017, XPLiu_2018,yao_degenerate_2019, ZhenlianS_2023},
and also attract some theoretic interests \cite{mingyuan_2016, shaoliang_2017, zhang_effective_2017, yao_strongly_2018, yao_zhang_2018}.
Some $g$-wave Feshbach resonances have also been studied in recent experiments \cite{zhang_transition_2021, ZhenlianS_2023}.

Spectroscopy is a powerful tool to investigate the rich many-body physics in ultracold quantum gases.
Various spectroscopies are used in ultracold quantum gas experiments, for example, RF spectroscopy, Bragg spectroscopy, and lattice modulation spectroscopy \cite{torma_physics_2016, haoyue_2024}.
The RF spectral response of unitary $^{6}\mathrm{Li}$ Fermi gas at different temperatures is observed in \cite{biswaroop_2019}.
RF response of impurities interacting with a quantum gas at finite temperature is investigated theoretically in \cite{weizhe_2020,weizhe_pra_2020}.
However, RF spectroscopy of high partial wave interacting quantum gases is rarely studied.

\begin{figure}[h]
  \centering
  \includegraphics[width=\linewidth]{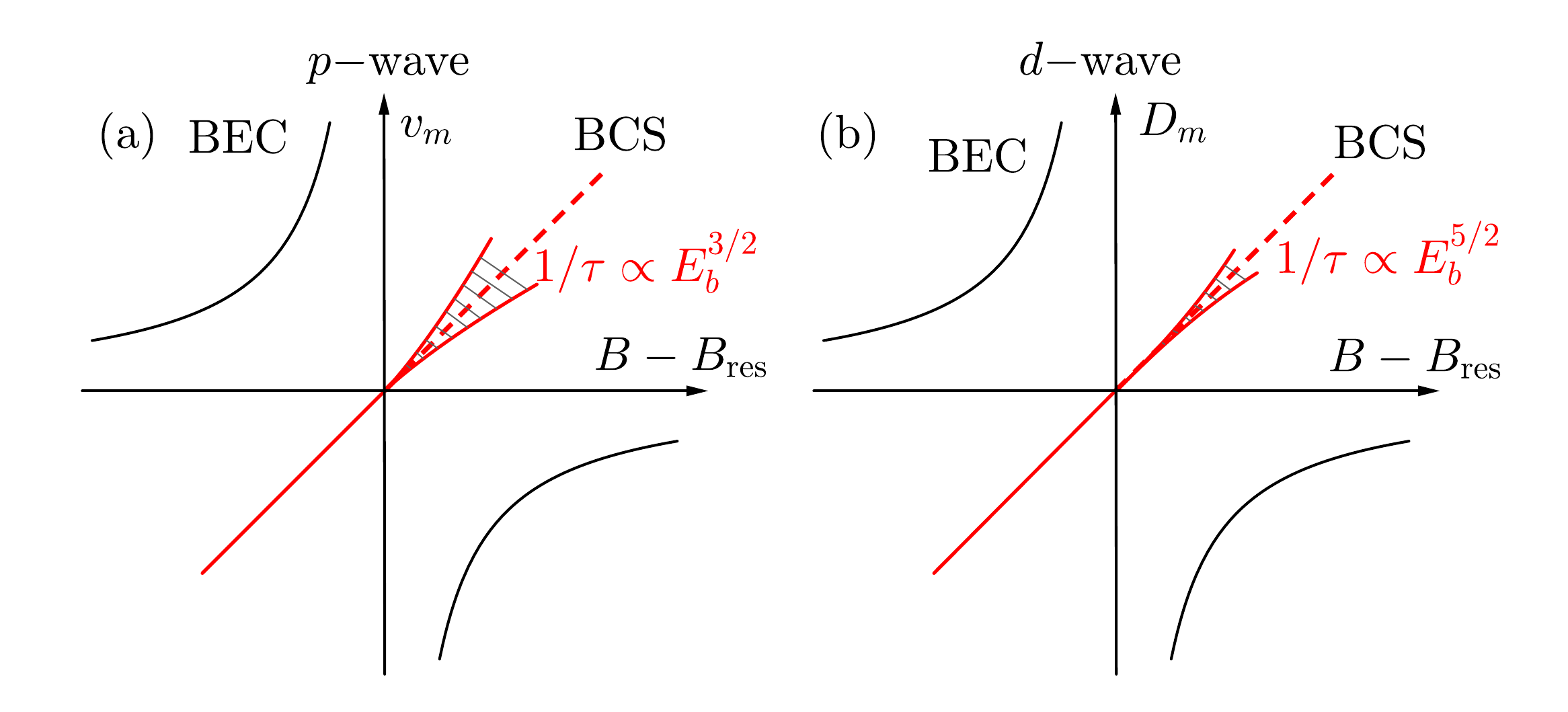}
  \caption{Schematic of different interaction regions.
  The black solid lines show the scattering volume $v_m$ (super-volume $D_m$) as a function of magnetic field $B$, and diverges at resonance magnetic field $B_{\mathrm{res}}$. The red solid lines indicate the shallow bound state on the BEC side. On the BCS side, the red dashed lines indicate the quasi-bound state with positive binding energy and finite lifetime $\tau$ represented by the shaded area.
  }
  \label{fig:schematic}
\end{figure}

Motivated by the recent experiments for high partial wave interacting quantum gases, we investigated the RF spectroscopy and momentum distribution of quantum Bose gases across $p$-wave and $d$-wave resonance in different interaction strength regions.
The low energy scattering amplitude can be characterized by the scattering volume $v_m$ (super-volume $D_m$) for $p$-wave ($d$-wave) interaction whose definition will be given in Sec. \ref{sec:model}. In experiments, $v_m$ ($D_m$) can be tuned from negative to positive infinity by tuning the magnetic field \cite{luciuk_evidence_2016,cui_2017,yao_degenerate_2019}, as illustrated schematically in Fig. \ref{fig:schematic}. When $v_m>0$ ($D_m>0$), there exists a shallow bound state with infinite lifetime, while for $v_m<0$ ($D_m<0$), there is a quasi-bound state with positive binding energy due to the centrifugal barrier.
As shown in Fig. \ref{fig:schematic}, this quasi-bound state has a finite lifetime $\tau$ proportional to $1/E_b^{3/2}$ for $p$-wave ($1/E_b^{5/2}$ for $d$-wave)  with $E_b$ being the binding energy. The inverse lifetime $1/\tau$ characterizes the coupling strength between the quasi-bound state and the scattering states which have important influences in the RF spectrum. In analogy to the BEC-BCS crossover of Fermions, we will call the $E_b<0$ region BEC side and the $E_b>0$ region BCS side for convenience in this article.

We first introduce our two-channel model in Sec. \ref{sec:model}.
The self-energy of single particle Green's function is obtained within ladder diagram approximation or the so-called Nozi\`{e}res and Schmitt-Rink (NSR) scheme \cite{nozieres_bose_1985,yao_strongly_2018}.
In Sec. \ref{sec:contact}, the contact density is also derived within NSR scheme through the adiabatic theorem \cite{yu_universal_2015,tan_energetics_2008,zhang_effective_2017}, and is compared with the results from high temperature virial expansion.
Next, in Sec. \ref{sec:spectral-function}, we present the results of single particle spectral function in which a clear anisotropic dependence in the momentum will be shown.
The momentum distributions, obtained by integrating the single particle spectral function over frequency, are analyzed in Sec. \ref{sec:momentum}.
Finally, results of RF spectrum are presented and an estimation of signal strength (in terms of the atom transfer rate to final state) under typical experimental parameters is provided in Sec. \ref{sec:rf}.

\section{Model}\label{sec:model}

\subsection{Hamiltonian}\label{sec:Ham}

For strongly interacting Bose atoms across a $p$-wave or $d$-wave FR,
the Hamiltonian can be written as the following two-channel model \cite{yao_strongly_2018},
\begin{align}
  &\hat{H}_0 = \sum_{\boldsymbol{k},\sigma=\sigma_{1,2}}\xi_{\boldsymbol{k}}\hat{a}_{\boldsymbol{k},\sigma}^{\dagger}
            \hat{a}_{\boldsymbol{k},\sigma}
           + \sum_{\boldsymbol{q},m}\left(\xi_{\boldsymbol{q},b} -
                \nu_m\right)\hat{b}_{\boldsymbol{q},m}^{\dagger} \hat{b}_{\boldsymbol{q}, m}\\
  \nonumber
 &+ \sqrt{\frac{4\pi}{V }} \sum_{\boldsymbol{k},\boldsymbol{q},m}\left[g_m k'^l Y_{l, m}(\hat{\boldsymbol{k}}')
   \hat{b}^{\dagger}_{\boldsymbol{q},m}\hat{a}_{-\boldsymbol{k}+\boldsymbol{q},\sigma_1}\hat{a}_{\boldsymbol{k},\sigma_2}
   + \mathrm{h.c.} \right],
\end{align}
where $\hat{a}_{\boldsymbol{k},\sigma}$ and $\hat{b}_{\boldsymbol{q}, m}$ are annihilation
operators for atoms and closed channel dimers and $\sigma$ labels the different hyperfine states.
The first and second terms correspond to the energy of free atoms and dimers.
$\xi_{\boldsymbol{k}} = \frac{\hbar^2k^2}{2M} - \mu$ is the kinetic energy of an atom measured from its chemical potential $\mu$.
$\xi_{\boldsymbol{q},b} = \frac{\hbar^2q^2}{4M} - 2\mu$ is the kinetic energy for a dimer measured from its chemical potential $2\mu$.
To simplify the notations, we will set $M =k_B=\hbar= 1$ ($k_B$ being the Boltzmann constant) throughout the rest part of this paper.
$\nu_m$ is the detuning for $m$ channel with $m$ standing for magnetic quantum number.
The last term in the Hamiltonian describes the coupling between the atom channel and the dimer channel with $l=1,2$ corresponding to $p$- and $d$-wave resonances respectively \cite{luciuk_evidence_2016,cui_2017,yao_degenerate_2019}. $m=0,\pm1,\cdots,\pm l$ represent the quantum number of $\hat{l}_z$.
$\boldsymbol{k}'= \boldsymbol{k}-\boldsymbol{q}/2$ is the relative momentum of two
incoming atoms that form the dimer, and $k' = |\boldsymbol{k}'|$, $\hat{\boldsymbol{k}}'=\boldsymbol{k}'/k'$.
$V$ is the volume of the system.
$g_m$ is the strength of coupling and $Y_{l, m}(\theta, \phi)$ are the spherical harmonic functions.

In this work, we consider a system very close to a certain $p$-wave or $d$-wave Feshbach resonance and far away from any other s-wave resonances. As a result, there are no low-energy $s$-wave bound states in the system under consideration. However, usually there will still be a small background $s$-wave scattering length $a_{bg}$ in this region. In most cases, $a_{bg}$ is on the same order of Van der Waals length such that $a_{bg}n^{1/3}\ll 1$ for typical atom density $n$. We expect that the main effect of such a small residue $s$-wave interaction is to induce a meanfield shift of the chemical potential $\delta\mu\sim4\pi\hbar^2 n a_{bg}/m$, which does not change any of our results qualitatively. We thus believe that it should be a good approximation to neglect the $s$-wave interaction in this paper. Because the Bosonic wave function must be symmetric under the exchange of two identical Bosons, $p$-wave interaction can only exist between two different hyperfine states.
As a result, for $p$-wave interaction we consider a two components Bose gas where the summation of $\sigma$ is over $\sigma_{1,2}=1,2$, while for $d$-wave interaction we consider a single component Bose gas with $\sigma_{1}=\sigma_{2}$ and the summation over $\sigma$ drops out.
In analogy to the Fermi gas, we will define a characteristic momentum unit
  $k_F$ as $k_F^3 = 3(6)\pi^{2}n_{\mathrm{total}}$
  for the two(single) component Bose gas, and a corresponding energy unit $E_F = k_F^2/2$. Here $n_{\mathrm{total}}$ is the total particle number density of Bose gas.

The bare parameters $\nu_m$ and $g_m$ are related to the low energy scattering parameters through a set of renormalization relations \cite{yao_strongly_2018,zhang_effective_2017,yao_zhang_2018}.
For $p$-wave interaction, we have
\begin{align}
  \frac{\nu_m}{g_m^2} =& \frac{1}{4\pi} v_{p,m}^{-1} - \frac{1}{V}\sum_{\boldsymbol{k}}1,\\
  \frac{1}{g_m^2} =& \frac{1}{4\pi}R_{p,m}^{-1} - \frac{1}{V}\sum_{\boldsymbol{k}} \frac{1}{k^2},
\end{align}
while for $d$-wave, we have
\begin{align}
  \frac{\nu_m}{g_m^2}  =&\frac{1}{4\pi}D_m^{-1}  - \frac{1}{V}\sum_{\boldsymbol{k}} k^2,\\
   \frac{1}{g_m^2}  =&\frac{1}{4\pi} v_{d,m}^{-1} - \frac{1}{V}\sum_{\boldsymbol{k}} 1, \\
   R_{d,m}^{-1} = & \frac{4\pi}{V}\sum_{\boldsymbol{k}}\frac{1}{k^2},\label{eq:d3}
\end{align}
where the meaning of physical parameters $D_m$, $v_{p(d),m}$ and $R_{p(d),m}$ will be clear in the two-particle vertex function presented in the next section.

\subsection{Self-energy}
\begin{figure}[h]
  \centering
  \includegraphics[width=\linewidth]{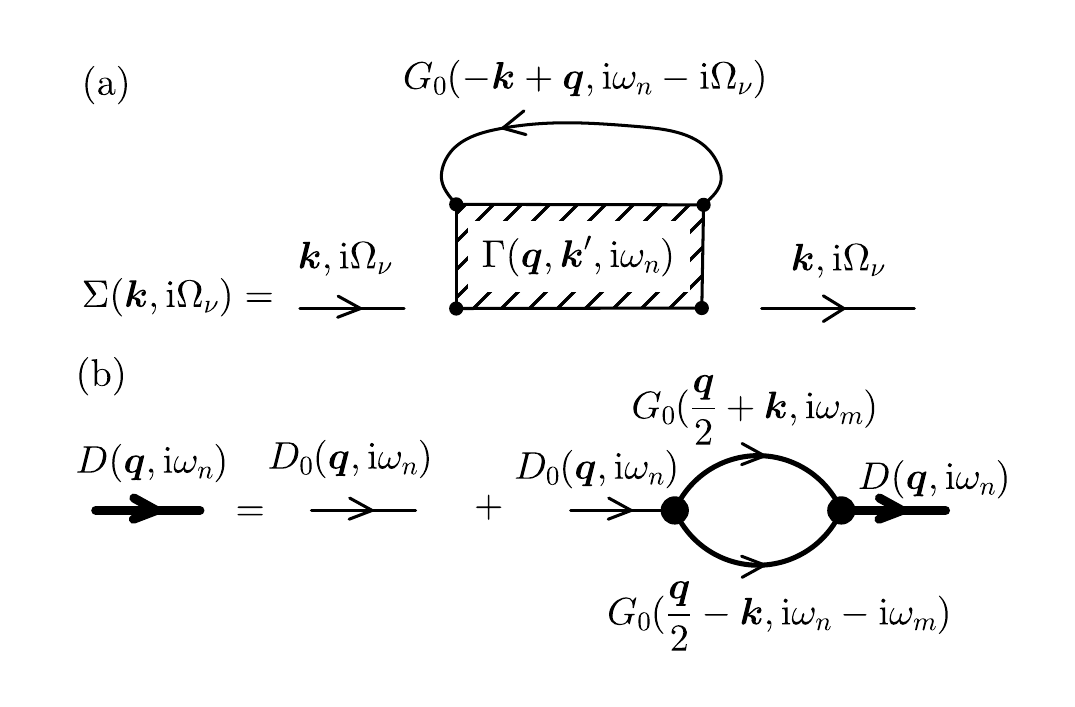}
  \caption{Feynman diagram for (a) the self-energy of atoms and (b) the Green's function of dimers.}
  \label{fig:selfenergy}
\end{figure}

The self-energy of single particle Green's function is obtained within the ladder diagram approximation as shown in Fig. \ref{fig:selfenergy} \cite{nozieres_bose_1985,yao_strongly_2018}:
\begin{align}
  \Sigma (\boldsymbol{k}, \mathrm{i}\Omega_{\nu}) = -\frac{1}{\beta} \sum_{\boldsymbol{q}}
  \sum_{\omega_n} \Gamma(\boldsymbol{q}, \boldsymbol{k}', \mathrm{i}\omega_n)
  G_0(-\boldsymbol{k} + \boldsymbol{q}, \mathrm{i}\omega_n - \mathrm{i}\Omega_{\nu}),
\end{align}
where $\Omega_{\nu} = 2\nu\pi/\beta$ and $\omega_n = 2n\pi/\beta$ are Bose matsubara frequencies
with $\beta = 1/T$ is the reciprocal of temperature. $G_0(-\boldsymbol{k} + \boldsymbol{q},
\mathrm{i}\omega_n - \mathrm{i}\Omega_{\nu}) =1/(\mathrm{i}\omega_n - \mathrm{i}\Omega_{\nu}-
\xi_{-\boldsymbol{k} + \boldsymbol{q}})$ is
Matsubara Green's function for free atoms
and $\Gamma(\boldsymbol{q}, \boldsymbol{k}', \mathrm{i}\omega_n)$ is the two-particle vertex function within ladder approximation \cite{strinati_bcsbec_2018},
\begin{align}
\Gamma(\boldsymbol{q}, \boldsymbol{k}', \mathrm{i}\omega_n)
   = \sum_m\frac{4\pi}{V } g_m^2 k'^{2l} \left|Y_{l, m}(\hat{\boldsymbol{k}'})\right|^2
  D(\boldsymbol{q}, \mathrm{i}\omega_n),
\end{align}
which describes the scattering between two atoms with the center of mass
momentum $\boldsymbol{q}$ and relative momentum $\boldsymbol{k}'$.
$D(\boldsymbol{q}, \mathrm{i}\omega_n)$ is the Matsubara Green's function of dimers
formed by two atoms, which can be calculated by the Dyson equation in Fig. \ref{fig:selfenergy}(b):
\begin{align}
  D(\boldsymbol{q}, \mathrm{i}\omega_n) = \frac{1}
  {\frac{1}{D_0(\boldsymbol{q}, \mathrm{i}\omega_n)} + g_m^2\chi_{\mathrm{pp}}(\boldsymbol{q}, \mathrm{i}\omega_n)},
\end{align}
where $D_0(\boldsymbol{q}, \mathrm{i}\omega_n) = 1/(\mathrm{i}\omega_n -\xi_{\boldsymbol{q}, b} +
\nu_m)$ is the free dimer propagator. $\chi_{\mathrm{pp}}$ is the particle-particle
bubble
\begin{align}
  \label{eq:chi-pp}
  &\chi_{\mathrm{pp}}(\boldsymbol{q}, \mathrm{i}\omega_n)
    =\frac{4\pi}{V} \sum_{\boldsymbol{k}}k^{2l}|Y_{l, m}(\hat{\boldsymbol{k}})|^2\\
  &\times\nonumber \frac{1}{\beta}\sum_{\omega_m}
  G_0(\frac{\boldsymbol{q}}{2} + \boldsymbol{k}, \mathrm{i}\omega_m )
  G_0(\frac{\boldsymbol{q}}{2} - \boldsymbol{k}, \mathrm{i}\omega_n - \mathrm{i}\omega_m).
\end{align}
The divergent summation over $\boldsymbol{k}$ can be renormalized by the renormalization relations
in Sec. \ref{sec:Ham}.
Therefore we can write the two-particle vertex function
in a renormalized form for each $m$ channel:
\begin{align}
  \label{eq:gamma-p}
  \Gamma^p_m(\boldsymbol{q}, \boldsymbol{k}', \mathrm{i}\omega_n)
  = \frac{4\pi \frac{4\pi}{V}k'^2|Y_{1, m}(\hat{\boldsymbol{k}}')|^{2}}
    {\frac{1}{v_{p,m}} + \frac{Z}{R_{p,m}} + 4\pi R^p_{\mathrm{pp}}},
\end{align}
\begin{align}
  \label{eq:gamma-d}
  \Gamma^d_m(\boldsymbol{q}, \boldsymbol{k}', \mathrm{i}\omega_n)
  = \frac{4\pi \frac{4\pi}{V}k'^4|Y_{2, m}(\hat{\boldsymbol{k}}')|^{2}}
  {\frac{1}{D_m} + \frac{Z}{v_{d,m}} + \frac{Z^2}{R_{d,m}} + 4\pi R^d_{\mathrm{pp}}},
\end{align}
where $Z = \mathrm{i}\omega_n - \frac{q^2}{4} + 2 \mu$ and the renormalized particle-particle bubble
\begin{align}
R^p_\mathrm{pp} = \chi_{\mathrm{pp}} - \frac{1}{V}\sum_{\boldsymbol{k}} 1
  - Z \frac{1}{V}\sum_{\boldsymbol{k}}\frac{1}{k^2},
\end{align}
\begin{align}
  R^d_{\mathrm{pp}} = \chi_{\mathrm{pp}} - \frac{1}{V}\sum_{\boldsymbol{k}}k^2
    - Z \frac{1}{V}\sum_{\boldsymbol{k}}1 - Z^2 \frac{1}{V}\sum_{\boldsymbol{k}} \frac{1}{k^2}.
\end{align}
The superscripts $p$ and $d$ stand for $p$-wave case and $d$-wave case correspondingly.

A main difference for high partial wave interaction comparing to $s$-wave interaction is the existence of centrifugal barrier, which makes the coupling between the true or quasi-bound states and the many-body medium very weak. This allows us to neglect the contribution from the branch cut of the two-particle vertex function $\Gamma^{p,d}_m$ and only include their pole structure.
Within this single pole approximation, the two-particle vertex function will take the following simple form
\begin{align}
\Gamma^p_m(\boldsymbol{q}, \boldsymbol{k}', \mathrm{i}\omega_n)
   =& \frac{4\pi R_{p,m}\frac{4\pi}{V} k'^{2} \left|Y_{1, m}(\hat{k'})\right|^2 }
          {\mathrm{i}\omega_n - \xi_{\boldsymbol{q}, b} - E_b^p},
\end{align}
\begin{align}
\Gamma^d_m(\boldsymbol{q}, \boldsymbol{k}', \mathrm{i}\omega_n)
   =& \frac{4\pi v_{d,m}\frac{4\pi}{V} k'^{4} \left|Y_{2, m}(\hat{k'})\right|^2 }
          {\mathrm{i}\omega_n - \xi_{\boldsymbol{q}, b} - E_b^d}.
\end{align}
where the molecular binding energies are given as $E_b^p = -R_{pm}/v_{pm}$ for $p$-wave and $E_b^{d} = - v_{dm}/D_m$ for $d$-wave.
We can see that the effective range $R_{p, m}$ (super volume $v_{d, m}$), which characterized the effective range of the interaction, appears in the numerator of two-particle vertex function $\Gamma^{p(d)}_m$.

After the standard frequency summation, the self-energy of single-particle Green's function within ladder diagram approximation is given as
\begin{align}
  \label{eq:selfenergy-p}
  \Sigma^p_m (\boldsymbol{k}, \mathrm{i}\Omega_{\nu})
  =& \sum_{\boldsymbol{q}} \frac{4\pi}{V}4\pi R_{p,m} k'^{2}
  \left|Y_{1, m}(\hat{\boldsymbol{k}}')\right|^2 \\ \nonumber
   &\cdot \frac{N_{\mathrm{B}}(\xi_{-\boldsymbol{k} + \boldsymbol{q}} ) -N_{\mathrm{B}}\left(\xi_{\boldsymbol{q}, b} +E^p_b\right)}
     {\mathrm{i}\Omega_\nu + \xi_{-\boldsymbol{k} + \boldsymbol{q}} - \xi_{\boldsymbol{q}, b} -E^p_b},
\end{align}
\begin{align}
  \label{eq:selfenergy-d}
  \Sigma^d_m (\boldsymbol{k}, \mathrm{i}\Omega_{\nu})
  =& \sum_{\boldsymbol{q}} \frac{4\pi}{V}4\pi v_{d,m} k'^{4}
  \left|Y_{2, m}(\hat{\boldsymbol{k}}')\right|^2 \\ \nonumber
   &\cdot \frac{N_{\mathrm{B}}(\xi_{-\boldsymbol{k} + \boldsymbol{q}} ) -N_{\mathrm{B}}\left(\xi_{\boldsymbol{q}, b} + E^d_b\right)}
     {\mathrm{i}\Omega_\nu + \xi_{-\boldsymbol{k} + \boldsymbol{q}} - \xi_{\boldsymbol{q}, b} - E^d_b},
\end{align}
where $N_{\mathrm{B}}(\xi) = 1/(e^{\beta\xi} - 1)$ is the Bose-Einstein distrubution function.

In experiments, high partial wave resonances are usually split due to anisotropy dipole-dipole interaction. According to \cite{yao_degenerate_2019}, $m=0$ is most important in the quantum degenerate regime. Therefore, we will only take the $m = 0$ channel into account and omit the subscript $m$ without ambiguity in the following calculations.

\section{equation of state and contact}\label{sec:contact}
In this section, we first determine the equation of state for our system. The particle density at fixed temperature and chemical potential can be derived through the thermodynamic potential, which within the ladder approximation is given as
\begin{align}\label{eq:thermodynamic-potential}
  \mathcal{F} = \mathcal{F}_0 + \mathcal{F}_{\mathrm{int}},
\end{align}
where $\mathcal{F}_0 = \sum_{\sigma}1/\beta\sum_{\boldsymbol{k}}\ln [1 - e^{-\beta\xi_{\boldsymbol{k}}}]$ is the contribution of free atoms.
$\mathcal{F}_{\mathrm{int}}$ is the contribution of pair fluctuations due to the interaction between atoms, which can be derived using the Matsubara Green's function of dimers \cite{nozieres_bose_1985,yao_strongly_2018},
\begin{align}
  \mathcal{F}_{\mathrm{int}} = \frac{1}{\pi}\sum_{\boldsymbol{q}}\int_{-\infty}^{\infty}
  \mathrm{d}\omega
  \frac{1}{e^{\beta\omega} - 1}\delta(\boldsymbol{q}, \omega)
\end{align}
where the phase $\delta (\boldsymbol{q}, \omega)$ is defined as
$\delta (\boldsymbol{q}, \omega)= \mathrm{Arg} \left[1/D(\boldsymbol{q}, \omega)\right]$.
Follow the single pole approximation of last Section,
$\delta$ can be written as,
\begin{align}
  \delta(\boldsymbol{q}, \omega) =\pi \Theta\left(\xi_{\boldsymbol{q}, b} + E_b - \omega\right),
\end{align}
where $\Theta(x)$ is the Heaviside step function which is one for positive $x$ and zero for negtive $x$.

The total density can be expressed as the derivative of thermodynamic potential,
\begin{align}
  n_{\mathrm{total}} = -\frac{1}{V}\frac{\partial\mathcal{F}}{\partial \mu}
  =& -\frac{1}{V}\left(\frac{\partial}{\partial \mu} \mathcal{F}_0 +\frac{\partial}{\partial \mu}  \mathcal{F}_{\mathrm{int}} \right) \\
  =& n_0 + n_{\mathrm{int}},
\end{align}
where $n_0$ is the contribution of free atoms thermodynamic potential $\mathcal{F}_0$.
$n_{\mathrm{int}}$ is the contribution from the pair fluctuation,
\begin{align}
  \label{eq:density}
  n_{\mathrm{int}} = -\frac{1}{V}\frac{\partial\mathcal{F}_{\mathrm{int}}}{\partial \mu}
 = \frac{1}{\pi^2} \int_0^{\infty}\mathrm{d}q\cdot
     \frac{q^{2}}{e^{\beta(\frac{q^2}{4} - 2\mu + E^p_b)} - 1}.
\end{align}

Besides the particle density, we can also get the contact from the thermodynamic potential, which is an important parameter for many body systems \cite{tan_large_2008, tan_generalized_2008,tan_energetics_2008}.
We can get the contact density for $p$-wave scattering volume $v_p$ and $d$-wave super-volume $D$ by adiabatic theorem \cite{yu_universal_2015,tan_energetics_2008,zhang_effective_2017}
\begin{align}\label{eq:contact-p}
  \mathcal{C}^p  \equiv& \left.- \frac{1}{V}\frac{\partial}{\partial v_p^{-1}}\mathcal{F}^p \right|_{T, \mu} \\
   =& \frac{R_p}{2\pi^2}\int_0^{\infty} \mathrm{d}q\cdot q^2 N_B(\xi_{\boldsymbol{q}, b} + E^p_b)
   = R_p \frac{n_{\mathrm{int}}^p}{2},\label{Cp}
\end{align}
\begin{align} \label{eq:contact-d}
  \mathcal{C}^d \equiv& \left.-\frac{1}{V} \frac{\partial}{\partial D^{-1}} \mathcal{F}^d \right|_{T, \mu} \\
   =& \frac{v_d}{2\pi^2} \int_0^{\infty} \mathrm{d}q\cdot q^2 N_B(\xi_{\boldsymbol{q}, b} + E^d_b)
  =v_d \frac{n^d_{\mathrm{int}}}{2}.\label{Cd}
\end{align}

The Bose-Einstein condensation will happen when temperature is sufficiently low. On the BCS side, the weakly interacting atoms will condensate at chemical potential $\mu=0$. On the BEC side, the critical temperature for the condensation of dimers can be determined by Thouless criterion.
The critical temperature are estimated in \cite{yao_strongly_2018}, which reach the asymptotic value $0.436 E_F$ in the BCS limit, while approaches $0.218E_F$ ($0.137E_F$) for $p$-wave ($d$-wave) in the BEC limit. As a result, at zero temperature one expects that the system should undergo a phase transition from atom BEC to dimmer (or atom-pair) BEC as the resonance is tuned from BCS side to BEC side. In this work, we will only consider the case above the critical temperature and leave the properties of low temperature superfluid phase for future studies.

\begin{figure}
  \includegraphics[width=\linewidth]{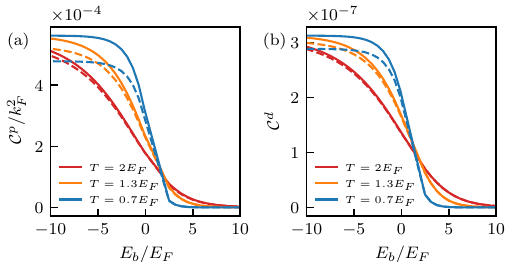}
\caption{The contact density as a function of binding energy at different temperatures for (a) $p$-wave and (b) $d$-wave.
  The solid lines are the results of Eq. \eqref{eq:contact-p} and Eq. \eqref{eq:contact-d}.
  The dashed lines are the results of virial expansion.
  The scattering parameters are fixed at $k_FR_p = k_F v_{d}^{1/3} =
  1/30$.}
\label{fig:contact}
\end{figure}

The contact for different interaction strengths at different temperatures above the critical temperature are shown in Fig. \ref{fig:contact}.
One can see that the $d$-wave contact is about three orders of magnitude smaller than that of $p$-wave contact(in units of  $E_F$),
which implies that the momentum distribution tails and the RF spectra tails for $d$-wave may also be much weaker than those of $p$-wave. We will see that this is indeed true as shown in Sec. \ref{sec:momentum} and \ref{sec:rf}.

Another interesting feature in Fig. \ref{fig:contact} is that, the temperature dependence of contact density is smoothly reversed when the interaction strength changes from BEC side to BCS side.
On the BEC side, the contact is mainly contributed from the dimmer state (true bound state) with negative binding energy, which tends to be broken when temperature becomes higher. As a result, the contact density becomes smaller when temperature increases. While on the BCS side, at higher temperature, more atoms are excited to positive energy that can be coupled to the quasi-bound state and thus contribute to a larger contact density.
This reversed temperature dependence of contact is also manifested in RF spectra as will be shown in Fig. \ref{fig:rf-temp}.


An alternative approach to calculate the contact density is the high temperature virial expansion, which becomes asymptotically exact in the limit $T\gg E_F$ \cite{huang_1987,liu_virial_2013}. In virial expansion, the thermodynamic potential is expanded as a Taylor series of fugacity $z = e^{\beta\mu}$,
\begin{align}\label{eq:cluster}
  \mathcal{F} = - \frac{1}{\beta}Q_1 \left[z + b_2 z^2 + \mathcal{O}(z^3)\right],
\end{align}
where $Q_1= 2 V/\lambda_{\mathrm{dB}}^3$ and $\lambda_{\mathrm{dB}}=\sqrt{2\pi\beta}$ is the thermal de Broglie wavelength.
The second virial coefficient $b_2$ can write as $b_2 = b_2^{(1)} + \Delta b_2$, where $b_2^{(1)}$ is the contribution of free atom thermodynamic potential $\mathcal{F}_0$. For $\Delta b_2$, we use the Beth-Uhlenbeck formalism second virial coefficient, which is expressed in terms of phase shifts of a two-body scattering problem \cite{beth_quantum_1937,liu_virial_2013,huang_1987}
\begin{align}\label{eq:b2}
  \frac{\Delta b_2}{\sqrt{2}} = \sum_i e^{-E^i_b/T} + \sum_l \frac{(2l + 1)}{\pi}
  \int_0^{\infty}\mathrm{d}k \frac{\mathrm{d}\delta_l}{\mathrm{d}k}
    e^{-\frac{\lambda_{\mathrm{dB}}^2k^2}{2\pi}},
\end{align}
where the first summation is over all the two-body bound states, which only exist on the BEC side in our system.
The phase shift in the second term is given by $k^3\cot\delta_p = -1/v_p - k^2/R_p$ and $k^5\cot\delta_d = -1/D - k^2/v_d- k^4/R_d$ \cite{yu_universal_2015,zhang_effective_2017}. As shown in Fig. \ref{fig:contact}, the result from ladder approximation agrees well with that from virial expansion at higher temperature which provides a non-trivial check for our method.

\section{Single particle spectral functon}\label{sec:spectral-function}
\begin{figure}[h]

  \includegraphics[width=1\linewidth]{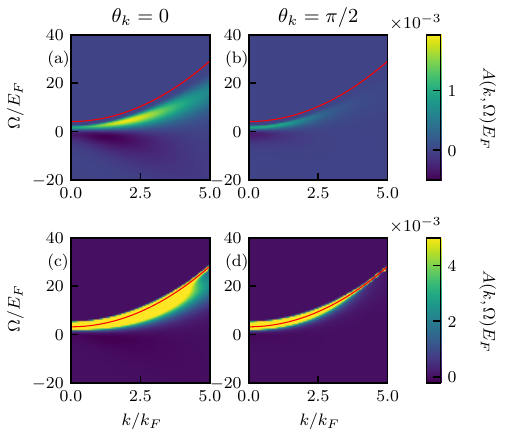}
  \caption{Single particle spectral function for $p$-wave in different momentum direction.
    (a) and (b) are with binding energy $E_b = -2 E_F$.
    (c) and (d) are with binding energy $E_b = 2 E_F$.
    The red line in (a) and (b) indicates the free atom contribution.
    The red line in (c) and (d) indicates the location of $\Omega = \xi_{\boldsymbol{k}}$.
    The results are calculated at temperature $T = 2E_F$.
    Effective range is $k_FR_p = 1/30$. In order to see the peak
    clearly, we set the color bar limit $5\times 10^{-3}$ in (c) and (d).
  }
  \label{fig:ako-p}
\end{figure}

\begin{figure}[h]

  \includegraphics[width=1\linewidth]{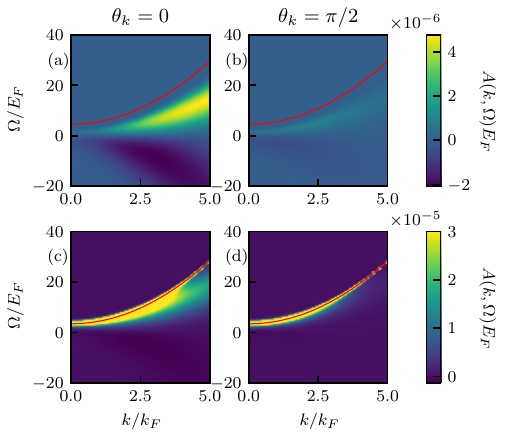}
  \caption{Single particle spectral function for $d$-wave in different momentum direction.
    (a) and (b) are with binding energy $E_b = -2 E_F$.
    (c) and (d) are with binding energy $E_b = 2 E_F$.
    The red line in (a) and (b) indicates the free atom contribution.
    The red line in (c) and (d) indicates the location of $\Omega = \xi_{\boldsymbol{k}}$.
    The results are calculated at temperature $T = 2E_F$.
    Effective range is $k_F v_d^{1/3} = 1/30$. In order to see the peak
    clearly, we set the color bar limit $3\times 10^{-5}$ in (c) and (d).
  }
  \label{fig:ako-d}
\end{figure}

In this section, we calculate the single particle spectral function, from which both the momentum distributions and RF spectrum will be derived in the following sections.
For the convenience of numerical calculation of the single particle spectral function and many other properties of our system,
we change the Matsubara Green's functions Eqs. \eqref{eq:selfenergy-p} and \eqref{eq:selfenergy-d} to retarded Green's functions by analytic continuation,
\begin{align}
 \Sigma(\boldsymbol{k}, \mathrm{i}\Omega_{\nu}) \rightarrow \Sigma(\boldsymbol{k}, \Omega + \mathrm{i}0^+).
\end{align}
The single particle spectral function is then written as \cite{mah00},
\begin{align}
  \label{eq:spectral-function}
  A(\boldsymbol{k}, \Omega) = \frac{-2 \mathrm{Im}\Sigma(\boldsymbol{k}, \Omega)}
  {\left[\Omega - \xi_{\boldsymbol{k}} - \mathrm{Re}\Sigma(\boldsymbol{k}, \Omega)\right]^2
   + \left[\mathrm{Im}\Sigma(\boldsymbol{k}, \Omega)\right]^2}.
\end{align}
In practice, we find that the term $\mathrm{Re}\Sigma(\boldsymbol{k}, \Omega)$ is much smaller than the kinetic energy term $\xi_{\boldsymbol{k}}$ and thus can be safely neglected in Eq. \eqref{eq:spectral-function} for both $p$- and $d$-wave resonance. The results are shown in Figs. \ref{fig:ako-p} and \ref{fig:ako-d}.

The single particle spectral function of our system mainly contains the following two features: (i) the single particle excitation peak $A_0$ centered at the free atom excitation energy $\Omega = \xi_{\boldsymbol{k}}$ whose lifetime is determined by $\mathrm{Im}\Sigma(\boldsymbol{k}, \Omega=\xi_{\boldsymbol{k}})$, and (ii) a threshold behavior from breaking a dimmer denoted by $A_{\mathrm{int}}$.

On the BEC side, $A_0$ is a delta peak with infinite lifetime,
\begin{align}
A_0(\boldsymbol{k}, \Omega) = 2\pi \delta(\Omega - \xi_{\boldsymbol{k}}),
\end{align}
indicated by red lines in Figs. \ref{fig:ako-p}(a), \ref{fig:ako-p}(b) and \ref{fig:ako-d}(a), \ref{fig:ako-d}(b). This delta peak results from the fact that the denominator of $\Sigma(\boldsymbol{k}, \Omega=\xi_{\boldsymbol{k}})$, $\xi_{\boldsymbol{k}} + \xi_{-\boldsymbol{k} + \boldsymbol{q}} - \xi_{\boldsymbol{q}, b} - E_b = (\boldsymbol{k}- \boldsymbol{q}/2)^2 -E_b$, is always positive when $E_b<0$, which implies $\mathrm{Im}\Sigma(\boldsymbol{k}, \Omega=\xi_{\boldsymbol{k}})=0$. Since the energy of an atom in the dimer is far away from the energy of a free atom, the spectral peak of the free atom and that of an atom in a dimer are well separated. This result suggests that there is no coupling between the dimmer state with negative binding energy and scattering continuum with positive energy at this level of approximation.

On the BCS side, there is only quasi-bound state with positive binding energy which has finite coupling with the scattering continuum. As a result, the peak from the single particle excitation now gets a finite lifetime and merges into the dimer excitation, as shown in   Figs. \ref{fig:ako-p}(c), \ref{fig:ako-p}(d) and \ref{fig:ako-d}(c), \ref{fig:ako-d}(d).

We can also see from Fig. \ref{fig:ako-p} and Fig. \ref{fig:ako-d} that the strength of the single particle spectral functions is very different at different momentum directions (here, we show the $\theta_k=0$ and $\theta_k=\pi/2$ direction), which reflect the anisotropic nature of an isolated $m=0$ FR of high partial wave interaction.

\section{Momentum distribution}\label{sec:momentum}
In this section, we calculate the momentum
distribution of atoms which can be expressed through the single particle spectral function obtained in Sec. \ref{sec:spectral-function},
\begin{align}
  \label{eq:momentum}
  N(\boldsymbol{k}) = \int_{-\infty}^{+\infty} \frac{\mathrm{d}\Omega}{2\pi} N_{\mathrm{B}}(\Omega)
  A(\boldsymbol{k}, \Omega).
\end{align}
In the limit $k= |\boldsymbol{k}|\rightarrow\infty$,
we obtain the following asymptotic
behavior for the momentum distribution:
\begin{align}
  \label{eq:m-t-p}
  N^p(\boldsymbol{k}) \approx \frac{16\pi^2}{k^2} \mathcal{C}^p \left|Y_{1,0}(\theta_k)\right|^2,
  \\
  \label{eq:m-t-d}
  N^d(\boldsymbol{k}) \approx 16\pi^2 \mathcal{C}^d \left|Y_{2,0}(\theta_k)\right|^2,
\end{align}
where $\mathcal{C}^p$ and $\mathcal{C}^d$ is given by Eqs. \eqref{Cp} and \eqref{Cd}.
The above results verify the well known $k^{-2}$ and $k^0$ tail as well as their relation with contacts for $p$-wave
and $d$-wave interaction proposed in \cite{yu_universal_2015, zhang_effective_2017}.

\begin{figure}[h]

\includegraphics[width=1\linewidth]{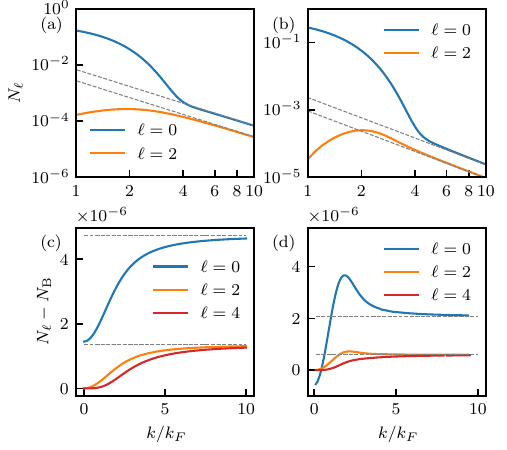}
\caption{
  The momentum distribution projection to Legendre functions.
  (a) and (b) for $p$-wave. (c) and (d) for $d$-wave.
 The analytic tail in Eq. \eqref{eq:m-t-p} and Eq. \eqref{eq:m-t-d} are indicated
 by gray dashed lines.
 (a) and (c) are with binding energy $E_b = - 2E_F$,
while (b) and (d) are with binding energy $E_b = 2E_F$.
The results are calculated at temperature $T = 2E_F$.
Effective range is $k_FR_p = k_Fv_d^{1/3} = 1/30$.
}
\label{fig:momentum-tail}
\end{figure}
The anisotropic nature of our system also manifests itself in the momentum distribution. Here we project the momentum distributions to Legendre function $P_{\ell}$,
\begin{align}
N_{\ell} (k) = \int_0^{\pi}\mathrm{d}\theta_k\cdot \sin\theta_k N(\boldsymbol{k}) P_{\ell}(\cos\theta_k).
\end{align}
The odd $\ell$ components are always zero which results from the spacial inversion symmetry. The even components are shown in Fig. \ref{fig:momentum-tail}.
The analytic momentum tail in Eqs. \eqref{eq:m-t-p} and \eqref{eq:m-t-d}
are indicated by the gray dashed lines, which agree well with the numerical results from Eq. \eqref{eq:momentum}.

\section{rf spectroscopy}\label{sec:rf}
Experimentally, the RF spectroscopy is measured by coupling the hyperfine state $\sigma_2$ to another hyperfine state
labeled as $\sigma_3$ by RF pulse. This coupling can be described by the following Hamiltonian
\begin{align}
 \hat{H}_L= & \sum_{\boldsymbol{k}}\left[
    \hat{a}_{\boldsymbol{k},\sigma_2}^{\dagger}\hat{a}_{\boldsymbol{k},\sigma_3}e^{\mathrm{i}\omega't}+ \mathrm{h.c.} \right],
\end{align}
where $\omega' = \nu_{\mathrm{rf}} + \mu - \mu_3$ with $\nu_{\mathrm{rf}}$ is the RF detuning.
Then the RF spectrum, which describes the atoms transferring rate from $\sigma_2$ to $\sigma_3$ state, can be expressed as a function of single particle spectral function \cite{torma_physics_2016},

\begin{align}
  I(\nu_{\mathrm{rf}}) =
   \frac{1}{2}\frac{1}{V}\sum_{\boldsymbol{k}}A(\boldsymbol{k}, \xi_{\boldsymbol{k}} - \nu_{\mathrm{rf}})
   \left[N_{\mathrm{B}}(\xi_{\boldsymbol{k}} - \nu_{\mathrm{rf}}) - N_{\mathrm{B}}(\xi_{\boldsymbol{k}, 3})  \right],
\end{align}
where $\xi_{\boldsymbol{k}, 3} = k^2/2 - \mu_3$.
In numerical calculation, we take the zero-density or vacuum limit where $\mu_3<0$ and $|\mu_3|\gg T$, such that the $\sigma_3$ state is nearly empty i.e. $N_{\mathrm{B}}(\xi_{\boldsymbol{k}, 3}) = 0$. Now the RF spectrum can be simplified as
\begin{align}
  \label{eq:rf}
I(\nu_{\mathrm{rf}}) = \frac{1}{2}\frac{1}{V}\sum_{\vec{k}} A(\boldsymbol{k}, \xi_{\boldsymbol{k}} -
  \nu_{\mathrm{rf}})N_{\mathrm{B}}(\xi_{\boldsymbol{k}} - \nu_{\mathrm{rf}}).
\end{align}

Following the analysis in Sec. \ref{sec:spectral-function}, the RF spectra in our system also contain two features: (i) a single particle peak $I_0(\nu_{\mathrm{rf}})$ inherited from $A_0$, corresponding to transferring a free atom from $\sigma_2$ to $\sigma_3$, and (ii) the part $I_{\mathrm{int}}(\nu_{\mathrm{rf}})$ corresponding to transferring an atom in bound state or quasi-bound state from $\sigma_2$ to $\sigma_3$.

\begin{figure}
  \centering
  \includegraphics[width=\linewidth]{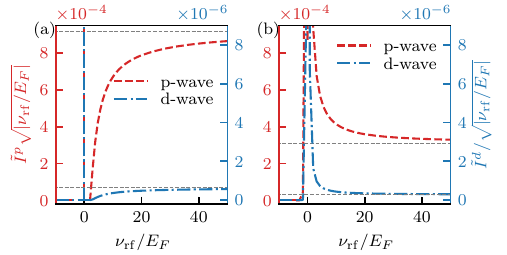}
  \caption{Dimensionless RF spectra of $p$-wave and $d$-wave at temperature $T = 2E_F$.
    (a) is with binding erergy $E_b = -2E_F$. (b) is with binding energy $E_b = 2E_F$.
    The effective range $k_FR_p = k_F v_d^{1/3} = 1/30$.
  Grey dashed lines show the tail in Eq. \eqref{eq:rf_tail-p} and Eq. \eqref{eq:rf_tail-d}
}
\label{fig:rf}
\end{figure}

On the BEC side, as shown in Fig. \ref{fig:rf}(a), $I_0(\nu_{\mathrm{rf}})$ is a delta peak located at $\nu_{\mathrm{rf}}=0$,
  \begin{align}
    I_0(\nu_{\mathrm{rf}}) = \pi \delta(\nu_{\mathrm{rf}}) \frac{1}{V}\sum_{\boldsymbol{k}} N_{\mathrm{B}} (\xi_{\boldsymbol{k}}),
    \end{align}
  which is well separated from $I_{\mathrm{int}}(\nu_{\mathrm{rf}})$. $I_{\mathrm{int}}(\nu_{\mathrm{rf}})$ has a threshold at $\nu_{\mathrm{rf}} = -E_b$, since the RF pulse must have enough energy to break the bound state in order to transfer an atom from $\sigma_2$ to $\sigma_3$ state.

On the BCS side, the weakly bounded pair has a positive binding energy. Therefore a red detuned RF pulse is enough to transfer the atoms in the weakly bounded pairs to the state $\sigma_3$. The free atom delta peak is expanded and merged into the  weakly bounded pairs, as shown in Fig. \ref{fig:rf}(b)

In the large frequency limit $\nu_{\mathrm{rf}}\rightarrow\infty$, we find the following asymptotic behavior from Eq. \eqref{eq:rf}
\begin{align}
  I^p(\nu_{\mathrm{rf}}) =& \frac{1}{\sqrt{\nu_{\mathrm{rf}}}} \mathcal{C}_p\label{eq:rf_tail-p}, \\
  I^d(\nu_{\mathrm{rf}}) =& \sqrt{\nu_{\mathrm{rf}}} \mathcal{C}_d\label{eq:rf_tail-d}.
\end{align}
This also confirms the universal large frequency tail proposed in \cite{yu_universal_2015, zhang_effective_2017}.
The grey dashed line in Fig. \ref{fig:rf}(b) indicated the contact density in Eq. \eqref{Cp} and Eq. \eqref{Cd} which agrees well with the numerical calculation of Eq. \eqref{eq:rf}.

\begin{figure}
  \includegraphics[width=\linewidth]{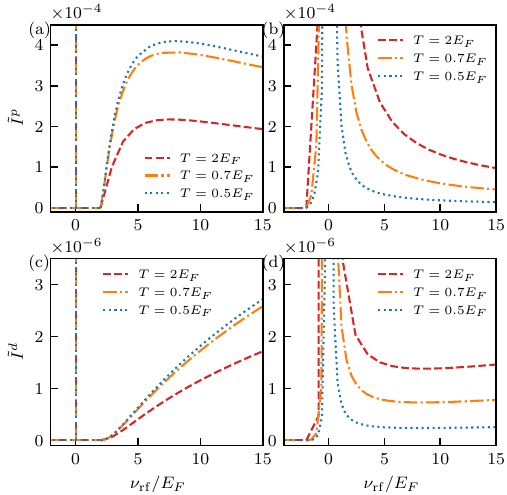}
  \caption{Dimensionless RF $\tilde{I}$ spectrum at different temperatures.
    (a) and (b) are for $p$-wave. (c) and (d) for $d$-wave.
    (a) and (c) are with binding energy $E_b = -2E_F$,
    while (b) and (d) are with binding energy $E_b = 2E_F$.
  Effective range is $k_FR_p = k_F v_d^{1/3} = 1/30$.}
\label{fig:rf-temp}
\end{figure}

Next, we investigated the RF spectra at different temperatures, as shown in Fig. \ref{fig:rf-temp}.
On the BEC side, the RF spectrum is more significant at lower temperatures, provided the system is in the normal phase. When the temperature increases, the number of dimers becomes smaller. Therefore the atom number transfer to state $\sigma_3$ from a dimer, i.e., the RF spectrum, becomes smaller. However, on the BCS side, because of a positive binding energy, the system tends to form more weakly bounded dimers when the temperature increases. Therefore the temperature dependence of the RF spectrum on the BCS side shows an opposite behavior. It is worth noting that this reversed temperature dependence was also found in the temperature dependence of the contact density in Fig. \ref{fig:contact}.

Finally, we give an experiment estimation. In experiments, the number of atoms transferred to the hyperfine state $\sigma_3$ in time $t$ obeys the sum rule \cite{luciuk_evidence_2016},
\begin{align}
  \label{eq:n-sumrule}
\int \frac{N_3(\nu_{\mathrm{rf}})}{t} \mathrm{d}\nu_{\mathrm{rf}} = \frac{\Omega_{\mathrm{Rb.}}^2}{2}\pi N,
\end{align}
where $\Omega_{\mathrm{Rb.}}$ is the Rabi frequency of RF pulse.
The sum rule of RF spectra in our work is
\begin{align}
  \label{eq:rf-sumrule}
\int \tilde{I}(\tilde{\nu}_{\mathrm{rf}}) \mathrm{d} \tilde{\nu}_{\mathrm{rf}} = \frac{1}{2},
\end{align}
where $\tilde{\nu}_{\mathrm{rf}} = \hbar\nu_{\mathrm{rf}}/ E_F$, and
\begin{align}
\tilde{I} = \frac{E_F}{\hbar} \frac{I(\nu_{\mathrm{rf}})}{2\pi}  \frac{V}{N},
\end{align}
is the dimensionless RF spectrum.

By comparing Eq. \eqref{eq:n-sumrule} and Eq. \eqref{eq:rf-sumrule}, we can express the atom number transferred to the hyperfine states $\sigma_3$ in time $t$ as
\begin{align}
N_3 =   \frac{\hbar}{E_F} \tilde{I} t\cdot  \Omega_{\mathrm{Rb.}}^2\pi N.
\end{align}
For a $10^5$ $^{41}\mathrm{K}$ atoms cloud with density $10^{13}\mathrm{cm}^{-3}$, the estimated $E^p_F = 2^{-2/3}E^d_F\approx 5475 h \mathrm{Hz} $ where $h$ is Planck constant. For a $0.16 \mathrm{ms}$ duration of RF pulse with Rabi frequency $\Omega_{\mathrm{Rb.}} = 30 \times 2\pi \mathrm{kHz}$,
the $p$-wave RF peak $\tilde{I}\approx 3\times 10^{-4}$ results in $ N_3\approx 1.6\times 10^4$, which is about $16\%$ of total atoms, transferred to $\sigma_3$ state.
For the $d$-wave RF peak, we find $\tilde{I} \approx 3\times 10^{-6}$ , and transferred atom number is $N_3\approx 98$, which is about $0.1\%$ of total atom number. It seems that the $p$-wave RF spectrum might be reachable for the present experiment techniques, while the observation of $d$-wave signal may be quite a challenge.

\section{summary}
In summary, we studied the $p$-wave and $d$-wave interacting Bose quantum gases in the normal phase using Green's function method in the spirit of ladder diagram approximation. Firstly, we derived the contact density, which agrees well with the results of virial expansion. Then, we calculated the single particle spectral function at different directions as well as the momentum distribution. The well known large $k$ tails for $p$- and $d$-wave interaction are confirmed. Finally, we obtain the RF spectrum and investigated the temperature dependence of the RF spectrum and discovered a reversed temperature dependence on the BCS and BEC side of FR. We also estimate the transferring rate of atom number under a typical experimental setup. The result for $p$-wave might be reachable, while for $d$-wave is quite a challenge.

\begin{acknowledgments}
  This work was supported by the National Key Research and Development Program of China (Grant No. 2022YFA1405301 and No. 2018YFA0306502), the National Natural Science Foundation of China (Grant No. 12022405 and No. 11774426), the Beijing Natural Science Foundation (Grant No. Z180013).
\end{acknowledgments}

\nocite{*}


\end{document}